\documentclass[letterpaper,numberedheadings, final]{aipproc}
\usepackage{dcolumn}
\usepackage{textcase}
\def\sb{\mbox{\rule{0pt}{11pt}}}

\def\al{\alpha}

\def\ga{\gamma}
\def\de{\delta}

\def\Up{\Upsilon}
\def\Dot{\!\cdot\!}
\layoutstyle{6x9}

\begin{document}

\title{Recent Developments in the Modeling of Heavy Quarkonia}

\classification{12.39.Pn,14.40.Lb,14.40.Nd}
\keywords      {}

\author{Stanley F. Radford}{
  address={Department of Physics, Marietta College, Marietta, OH 45750}
}

\author{Wayne W. Repko}{
  address={Department of Physics and Astronomy, Michigan State University, East Lansing, MI 48824}
}

\begin{abstract}
We examine the spectra and radiative decays of the $c\bar{c}$ and $b\bar{b}$ systems using a model which incorporates the complete one-loop spin-dependent perturbative QCD short distance potential, a linear confining term including (spin-dependent) relativistic corrections to order $v^2/c^2$, and a fully relativistic treatment of the kinetic energy. We compare the predictions of this model to experiments, including states and decays recently measured at Belle, BaBar, CLEO and CDF. 
\end{abstract}

\maketitle

%%%%%%%%%%%%%%%%%%%%%%%%%%%%%%%%%%%%%%%%%%%%
%% MAINMATTER
%%%%%%%%%%%%%%%%%%%%%%%%%%%%%%%%%%%%%%%%%%%%

\section{Introduction and the Potential Model}

Over the past 25+ years, potential models have proven valuable in analyzing the spectra and characteristics of heavy quarkonium systems \cite{prs,schnit,egkly,grr,gjrs,elq,bg}. Motivation for revisiting the potential model interpretation of the $c\bar{c}$ and $b\bar{b}$ systems is provided by recent experimental results:
\begin{itemize}
\item The discovery of several expected states in the charmonium spectrum ($\eta'_C$ and $h_C$)
\item The discovery of new states [$X(3872)$, $X(3943)$], which could be a interpreted as above threshold charmonium levels
\item The discovery of the $1^3D_2$ state of the upsilon system
\item The determination of various $E1$ widths for $c\bar{c}$ and $b\bar{b}$.
\end{itemize}
Our purpose here is to examine to what extent a semi-relativistic potential model which includes all $v^2/c^2$ and one-loop QCD corrections can fit the below threshold $c\bar{c}$ and $b\bar{b}$ data and accommodate the new above threshold states.

%\section{The Potential Model}
%
In our analysis, we use a semi-relativistic Hamiltonian of the form
\begin{eqnarray*}
H&=&2\sqrt{\vec{p}^2+m^2}+Ar-\frac{4\al_S}{3r}\left(1+\frac{\al_S}{3\pi}(12-n_f) (\ln(\mu r)+\ga_E)\right) +V_S+V_L \\
&=&H_0+V_S+V_L\,,
\end{eqnarray*}
where $\mu$ is the renormalization scale, $V_L$ contains the $v^2/c^2$ corrections to the confining potential and the short distance potential is $V_S=V_{HF}+V_{LS}+V_T+V_{SI}$, with   
\begin{eqnarray*}
V_{HF}\!\!\!\!\!&=&\!\!\!\frac{32\pi\al_S\vec{S}_1\Dot\vec{S}_2}{9m^2}\left\{\left[1-\frac{\al_S}{12\pi} (26+9\ln\,2)\right]\de(\vec{r})\right. \\ \nonumber
& &\left.-\frac{\al_S}{24\pi^2}(33-2n_f)\nabla^2\left[\frac{\ln\,\mu r+\ga_E}{r}\right]+\frac{21\al_S}{16\pi^2}\nabla^2\left[\frac{\ln\, mr+\ga_E}{r}\right]\right\} \\
V_{LS} &=&\!\!\!\frac{2\al_S\vec{L}\Dot\vec{S}}{m^2r^3}\!\left\{\!\!1\!-\! \frac{\al_S}{6\pi}\left[\frac{11}{3}-(33-2n_f)\ln\mu r+\!12\ln mr-(21\!-2n_f)(\ga_E-1)\!\right]\!\!\right\}\\ 
V_{T\;} &=&\!\!\!\frac{4\al_S(3\vec{S_1}\Dot\hat{r}\vec{S_2}\Dot\hat{r}-\vec{S_1}\Dot\vec{S_2})} {3m^2r^3}\left\{1+\frac{\al_S}{6\pi}\left[8+(33-2n_f)\left(\ln\mu r+\ga_E-\frac{4}{3}\right)\right.\right.\\ \nonumber
& &\left.\left.-18\left(\ln mr+\ga_E-\frac{4}{3}\right)\right]\right\} \\
V_{SI}\!\!\!&=&\!\!\!\frac{4\pi\al_S}{3m^2}\!\left\{\!\!\left[1\!-\!\frac{\al_S}{2\pi}(1+\ln2)\right] \!\!\de(\vec{r})-\!\frac{\al_S}{24\pi^2}(33\!-\!2n_f)\nabla^2\!\!\left[\frac{\ln\,\mu r+\ga_E}{r}\right]\!-\!\frac{7\al_Sm}{6\pi r^2}\!\!\right\} \\ \nonumber
%& &\left.-\frac{7\al_Sm}{6\pi r^2}\right\}\,.
\end{eqnarray*}
Using the variational procedure described in Ref.\,\cite{DPF2004}, we fit the experimental data for the the charm and $\Up$ systems by varying the parameters $A,\al_S,m,\mu$ and $f_V$, the fraction of vector coupling in the scalar-vector mixture of the confining potential, to find a minimum in $\chi^2$. This was done in two ways: first by treating $V_L+V_S$ as a perturbation and second by treating the entire Hamiltonian non-perturbatively. The results are shown in Table\,\ref{params}.
\begin{table}[h]
\centering
\begin{tabular}{lcccc}
\hline     
  & $c\bar{c}$\, Pert  & $c\bar{c}$\, Non-pert &
    $b\bar{b}$\, Pert  & $b\bar{b}$\, Non-pert \\
\hline
\sb$A$ (GeV$^2$) & 0.168   & 0.175   & 0.170    & 0.186 \\
\hline
\sb $\al_S$      & 0.331   & 0.361   & 0.297    & 0.299 \\
\hline
\sb $m_q$ (GeV)  & 1.41    & 1.49    & 5.14     & 6.33  \\
\hline
\sb $\mu$ (GeV)  & 2.32    & 1.07    & 4.79     & 3.61  \\
\hline
\sb $f_V$        & 0.00    & 0.18    & 0.00     & 0.09  \\
\hline
\end{tabular}
\caption{Fitted Parameters \label{params}}
\end{table}

\section{Results and Conclusions}
Our results\footnote{See: http://www.panic05.lanl.gov/sessions\_by\_date.php\#sessions3} for the fit to the $c\bar{c}$ spectrum and the predicted $E1$ transition rates from the resulting wave functions are comparable to recent results for charmonium \cite{elq,bg}, with the non-perturbative treatment yielding the best fit. The non-perturbative results$^1$ for the $b\bar{b}$ spectrum and decays are quite reasonable, though not as good as those from the perturbative treatment. In Table\,\ref{Upsspec}, we show the fit to the $b\bar{b}$ spectrum for the case of the perturbative treatment of $V_L+V_S$, and in Table\,\ref{E1}, we show our predictions for the observed $E1$ transitions and the for the $E1$ decays associated with the $\Up(1^3D_2)$. 

Aside from the above threshold states in $c\bar{c}$, where mixing as well as continuum effects must be included to describe the $X(3872)$ and the $X(3943)$, both treatments of $c\bar{c}$ and $b\bar{b}$ yield very good overall fits. It is striking that for both systems the perturbative fits require the confining terms to be pure scalar, while the non-perturbative fits require a small amount of vector exchange. 
\begin{table}[h]
\centering
\begin{tabular}{cllcll}
\hline 
\sb          & Pert   & Expt            &          & Pert  &  Expt      \\
\hline
\sb$\eta_b(1S)$ & 9411.6 & $9300\pm 28$   & $\eta_b(3S)$ & 10339.5   &  \\
\hline
\sb$\Up(1S)$     & 9459.5 &$9460.3\pm 0.26$& $\Up(3S)$    & 10359.5  & $10355.2\pm 0.5$ \\
\hline
\sb$1\chi_{b0}$   & 9862.5 & $9859.44\pm 0.52$  & $3\chi_{b0}$& 10511.6 & \\
\hline
\sb$1\chi_{b1}$   & 9893.2 & $9892.78\pm 0.40$  & $3\chi_{b1}$& 10534.5 & \\
\hline
\sb1$\chi_{b2}$   & 9914.0 & $9912.21\pm 0.17$  & 3$\chi_{b2}$& 10549.8 & \\
\hline
\sb$1h_b$         & 9902.1 &                    & $3h_b$      & 10540.9 & \\
\hline
\sb $\eta_b(2S)$  & 9996.5 &                    & $1^3D_1$    & 10149.8. & \\
\hline
\sb$\Up(2S)$ & 10020.9 &$10023.26\pm 0.31$ & $1^3D_2$        & 10157.6   &  $10161.1\pm 1.7$ \\
\hline
\sb$2\chi_{b0}$   & 10228.9 & $10232.5\pm 0.6$  & $1^3D_3$   & 10163.5  &  \\
\hline
\sb$2\chi_{b1}$   & 10254.0 & $10255.46\pm 0.55$& $1^1D_2$   & 10158.9  & \\
\hline
\sb$2\chi_{b2}$   & 10270.8 & $10268.65\pm 0.55$&            &          & \\
\hline
\sb$2h_b$         & 10261.1 &                   &            &          & \\
\hline
\end{tabular}
\caption{\label{Upsspec}}
\end{table}

\begin{table}[h]
\centering
\begin{tabular}{lcclcc}
\hline
\sb $\Gamma_{\ga}(E1)$\,(keV)& Pert & Expt &$\Gamma_{\ga}(E1)$\,(keV)& Pert & Expt \\ 
\hline
\sb$\Upsilon(2S)\to\ga\,1\chi_{b0}$ & 1.12 & $1.16\pm 0.15$
  &$\Upsilon(3S)\to\ga\,2\chi_{b0}$ & 1.64 & $1.30\pm 0.20$ \\
\hline
\sb$\Upsilon(2S)\to\ga\,1\chi_{b1}$ & 1.79 & $2.11\pm 0.20$   
  &$\Upsilon(3S)\to\ga\,2\chi_{b1}$ & 2.61 & $2.78\pm 0.43$ \\
\hline
\sb$\Upsilon(2S)\to\ga\,1\chi_{b2}$ & 1.76 & $2.19\pm 0.20$ 
  &$\Upsilon(3S)\to\ga\,2\chi_{b2}$ & 2.59 & $2.89\pm 0.50$\\
\hline
\sb$2\chi_{b1}\to\ga\,\Upsilon(1^3D_2)$ & 1.47 & 
  &$2\chi_{b2}\to\ga\,\Upsilon(1^3D_2)$ & 0.47 &  \\
\hline
\sb$\Upsilon(1^3D_2)\to\ga\,1\chi_{b1}$ & 19.7 &   
  &$\Upsilon(1^3D_2)\to\ga\,1\chi_{b2}$ & 5.16 &  \\
\hline
\end{tabular}
\caption{\label{E1}}
\end{table}

%%%%%%%%%%%%%%%%%%%%%%%%%%%%%%%%%%%%%%%%%%%%%%%%
%% BACKMATTER
%%%%%%%%%%%%%%%%%%%%%%%%%%%%%%%%%%%%%%%%%%%%%%%%

\begin{theacknowledgments}
This research was supported in part by the National Science Foundation under Grant PHY-0244789.
\end{theacknowledgments}

\end{document}